# Understanding the Interaction between Energetic Ions and Freestanding Graphene towards Practical 2D Perforation

*Jakob Buchheim[†], Roman M. Wyss[†], Ivan Shorubalko[‡,\*], Hyung Gyu Park[†,\*]*

[†]Nanoscience for Energy Technology and Sustainability, Department of Mechanical and Process Engineering, Eidgenössische Technische Hochschule (ETH) Zürich, Tannenstrasse 3, CH-8092 Zürich, Switzerland

[‡] Laboratory for Reliability Science and Technology, Empa (Swiss Federal Laboratories for Materials Science and Technology), Überlandstrasse 129, CH-8600 Dübendorf, Switzerland

[\*] To whom correspondence should be addressed. E-mail: ivan.shorubalko@empa.ch (IS), parkh@ethz.ch (HGP).

ABSTRACT:

We report experimentally and theoretically the behavior of freestanding graphene subject to bombardment of energetic ions, investigating the ability of large-scale patterning of freestanding graphene with nanometer sized features by focused ion beam technology. A precise control over the $He^+$ and $Ga^+$ irradiation offered by focused ion beam techniques enables to investigate the interaction of the energetic particles and graphene suspended with no support and allows determining sputter yields of the 2D lattice. We find strong dependency of the 2D sputter yield on the species and kinetic energy of the incident ion beams. Freestanding graphene shows material semi-transparency to $He^+$ at high energies (10-30 keV) allowing the passage of >97% $He^+$ particles without creating destructive lattice vacancy. Large $Ga^+$ ions (5-30 keV), in contrast, collide far more often with the graphene lattice to impart significantly higher sputter yield of ~50%. Binary collision theory applied to monolayer and few-layer graphene can successfully elucidate this collision mechanism, in great agreement with experiments. Raman spectroscopy analysis corroborates the passage of a large fraction of $He^+$ ions across graphene without much damaging the lattice whereas several colliding ions create single vacancy defects. Physical understanding of the interaction between energetic particles and suspended graphene can practically lead to reproducible and efficient pattern generation of unprecedentedly small features on 2D materials by design, manifested by our perforation of sub-5-nm pore arrays. This capability of nanometer-scale precision patterning of freestanding 2D lattices shows practical applicability of the focused ion beam technology to 2D material processing for device fabrication and integration.

Discovery of isolated and stable graphene has launched a new research field to explore a variety of unforeseen properties of this two-dimensional (2D) material.[1] In particular graphene has drawn significant attention by showing extraordinary mechanical strength,[2] great electrical[3] and thermal conductivities,[4] and virtually uninhibited transmission of light[5] yet hermetic sealing against material permeation.[6] Potential graphene-based technology is proposed for various applications including ultimately permeable membranes[7] and flexible electronics,[8] of which embodiment can be propelled by large-scale synthesis capabilities such as chemical vapor deposition (CVD).[9, 10] Device integration of graphene to exploit its unique properties, on the other hand, will require selective patterning via etching or crystallographic modification through exposure to plasma[11] or energetic ions.[12] The evolution of the electrical properties and quality of supported or sandwiched graphene subject to ion irradiation has been investigated,[13-18] all assuring that graphene can be patterned by energetic ion irradiation. Nevertheless, it is the presence of a support structure that obscures a mechanistic understanding of the effects of ion bombardment on a 2D lattice of graphene because the bulk sputtering mechanism of the support[19] perplexes the otherwise clearly observable 2D sputtering mechanism. For example, effects of the secondary cascade interaction of bombarding particles with the support layer, such as ion implantation and substrate swelling,[13, 15, 20] can influence experimental results significantly, eventually hampering the extraction of 2D sputtering mechanism. Freestanding graphene has been patterned using transmission electron microscopy where pores[21] and vacancies[22] can be created with very high precision but limited in scale due to immense irradiation dose required. Nanometer-sized feature formation in graphene has been enabled by block-copolymer self-assembly[23] as well as by strain assisted metal intercalcination,[24] both limited to feature sizes of ~20 nm. In the pattern-size regime other than 20 nm, a focused ion beam milling (FIB) process of freestanding graphene can be a promising technology for practical applications at intermediate scales. Recently, the possibility of pattern generation in freestanding graphene by ion irradiation has been demonstrated by the formation of nanoscale pores,[7, 25] nanoribbons[26] or other geometries,[27] though these studies lack a mechanistic understanding of the energetic-ion–graphene interaction. More insight about the interaction has been obtained by a few theoretical investigations using Monte Carlo simulation.[17, 18, 28] Despite a growing understanding of the 2D sputtering mechanism including graphene amorphization upon $Ga^+$ irradiation,[29] there is few report of combined experimental and theoretical investigation about the fate of freestanding graphene layers subject to ion bombardment of various ion species and energies, hampering exploitation of the advanced manufacturing capability of the ion bombardment on freestanding graphene.

Here, we investigate experimentally and theoretically the perforation physics of unsupported, freestanding graphene by energetic ion bombardment. We differentiate the interaction schemes between energetic ions and graphene layers. Depending on the physical dimension and kinetic energy, incident accelerated ions can either penetrate the freestanding graphene or collide with the atoms of the 2D crystal to produce various vacancies for sputtering. We rationalize these interactions by assuming a binary collision process between incident ion and the graphene lattice. In great agreement with experimental results, our model successfully elucidates the ion interaction mechanism with freestanding graphene. According to the results, it is the graphene lattice spacing that can, on a certain condition, render this 2D material semi-transparent to incident energetic ions. Our understanding of the ion-carbon interaction enables to create vacancies and patterns on freestanding graphene by design, leading to an establishment of the reproducible and scalable perforation method for graphene membranes such as sub-3-nm and sub-4-nm pore arrays using $He^+$ and $Ga^+$ FIB, respectively.

RESULTS AND DISCUSSION:

High degree of control of three synthesis steps for graphene device fabrication (CVD graphene growth, transfer, and subsequent annealing process) yielded ultraclean, freestanding graphene samples (**Figure 1a**) showing very few graphene wrinkles and sparse contamination sites. These samples transferred to the FIB chamber were irradiated with energetic ions. When exposed to $He^+$ irradiation, graphene was resistant to a high dose of ions bombarding the monolayer. We could take images of freestanding monolayer graphene repeatedly at a standard imaging dose ($\sigma_{He^+} \approx 10^{18}$ m$^{-2}$) of helium ion microscope (HIM) without significant damage to samples, in agreement with pervious findings.[30, 31] The power of HIM and the cleanliness of the graphene samples allow to clearly distinguish the layer numbers of freestanding multilayer graphene (**Figure 1b**). However, in the same experiment on a FIB system using $Ga^+$ ions, we found out that the freestanding graphene quickly deteriorated and was etched away. These findings indicate that the interaction between $Ga^+$ and carbon in the graphene lattice is more destructive than that between $He^+$ and graphene; namely, a $Ga^+$ ion has higher probability to chop off carbon atoms from the 2D lattice than the a $He^+$ ion does. To precisely quantify the difference in C atom removal, the sputter yield is determined by patterning 200 nm circular features into monolayer graphene. The average number of vacancies produced per ion bombardment, or a sputter yield $\gamma$, shows significantly higher value for $Ga^+$ (**Figure 2a**). For instance, the sputter yield of freestanding graphene upon 30-keV $Ga^+$ bombardment is about 47% ($\gamma_{Ga^+} \approx 47\%$), which confirms qualitative findings of previous reports for $Ga^+$-based graphene sputtering.[27, 32] This corresponds to a total ion dose necessary to create a pattern in monolayer graphene by 30-keV $Ga^+$ ions of ~8.1 × 10$^{19}$ m$^{-2}$ (**Figure 2b**) which is very much in line with a recent report[29] of 9.5 × 10$^{19}$ m$^{-2}$ for $Ga^+$ at 35-keV. Furthermore, in our experiments $\gamma$ of freestanding monolayer graphene showed clear dependency on the energy of the incident particles. For Ga$^+$ the reduction of the accelerating voltage from 30 kV to 5 kV increases the sputter yield from $\gamma_{Ga^+} \approx 47\%$ to $81\%$ (**Figure 2a**), in line with predictions of Monte Carlo molecular dynamics simulations.[17, 18] Interestingly, these $\gamma_{Ga^+}$ values of monolayer graphene are significantly smaller than those of its 3D counterpart: reportedly 120-270% depending on carbon allotropes.[32, 33]

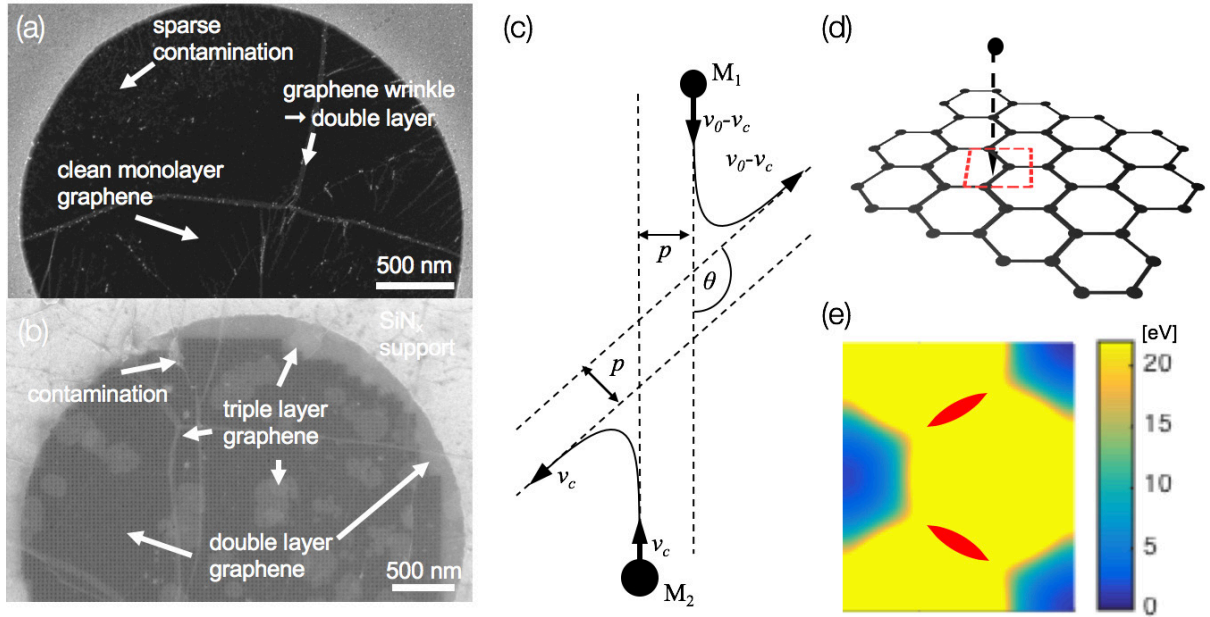

**Figure 1.** (**a**) SEM micrograph of freestanding clean monolayer CVD graphene after transfer on porous SiN$_x$ substrate. Brighter lines show graphene wrinkles (double layer) and small bright dots are sparse contaminates of the graphene. (**b**) Helium ion micrograph of He$^+$ patterned freestanding double layer graphene membrane. Local number variation of graphene layers can be clearly distinguished by pronounced brightness changes. (**c**) Binary collision model illustrated in a center-of-mass reference frame moving at a speed of $v_c$, where collision parameters $p$ and $\theta$ are the shortest projected distance and the scattering angle between two colliding particles, respectively (following reference[19]). (**d**) Schematic of an ion bombardment process of freestanding monolayer graphene. A red dashed box illustrates a unit cell of graphene. (**e**) A contour of superimposed energy from 15-keV Ga$^+$ to carbon atoms in a graphene unit cell, calculated from the binary collision model. Red shaded area depicts area of double vacancy production.

For He$^+$ FIB the sputtering yield is about two-order-of-magnitude lower ($\gamma_{He^+} \approx 0.7\%$) than for Ga$^+$ FIB. Therefore, pattern generation with 30-keV He$^+$ ions requires significantly higher dose of ~5.7 × 10$^{21}$ m$^{-2}$ (**Figure 2b**). Reaffirming the mechanism of HIM imaging,[30, 31] our finding sheds a renewed light on the possibility that, not only to proton,[34, 35] graphene can be nearly transparent to energetic He$^+$ ions as shown theoretically.[17] For example, at a kinetic energy of 30 keV approximately 99% of He$^+$ ions can penetrate through the monolayer graphene with statistically sputtering little or no carbon atom from the lattice, reminiscent of the photon and proton transmission. The material transparency of graphene is slightly reduced at lower He$^+$ acceleration voltages ($\gamma_{He^+, 10\,keV} \approx 2.4\%$, **Figure 2a**) yet only to ~97.6%. Both observations of the material transmission of H$^+$ [34, 35] and He$^+$ through the defect-free graphene lattice offers a new insight on the ability of graphene as the barrier material. In the case of increased particle energy or commensurable particle size, graphene crystals in their freestanding state (let alone crystallographic defects) indeed allow the permeation of small atoms through the lattice.

The interaction between energetic ions and carbon atoms in the graphene lattice can be elucidated further by considering the classic binary collision theory.[19, 36] This theory assumes an interception of two particle trajectories where the energetic ion at velocity $v_0$ is colliding with a carbon atom at velocity $v_c$ (**Figure 1c**). Depending on the minimum projected distance between two particle trajectories, $p$, the nuclei start to repel each other to avoid the overlap of the coulombic potential of the nuclei. The scatter angle, $\theta$, can be calculated using eq. (1):

$$\theta(p) = \pi - 2 \int_{r_{min}}^{\infty} r^{-2}(1 - V(r)/E_C - p^2/r^2)^{-1/2} dr, \qquad (1)$$

where $E_C = E_0 M_2 / (M_1 + M_2)$ denotes the collision energy with $E_0$ being the acceleration energy of the ion, $r$ the ion-to-atom center-of-mass distance, and $V(r)$ the Ziegler-Biersack-Littmark interatomic potential between ion and atom (**Figure 1c**). A good approximation for the interatomic potential $V(r) = \frac{Z_1 Z_2 e^2}{r} \Phi(r/a)$ is given by a repulsive coulombic potential created by the charges carried by the two nuclei corrected by a universal fit function $\Phi(r/a) = \sum_{i=1}^{4} C_i e^{k_i r/a}$ accounting for the electron cloud screening effects. The fit function depends on the distance of the scattering particles, $r$, the screening length, $a = 0.8854 a_0 (Z_1^{0.23} + Z_2^{0.23})^{-1}$, calculated by use of the Bohr radius, $a_0$, the two atomic charges $Z_1$ and $Z_2$ and has 8 parameters $C_1$= 0.1818, $C_2$= 0.5099, $C_3$= 0.2802, $C_4$= 0.02811 and $k_1$= -3.2, $k_2$= -0.9423, $k_3$= -0.4029, $k_4$= -0.2016.[19] It is valid for collisions with kinetic energies higher than a few 100 eV, where interatomic interactions are governed primarily by repulsive nuclei such that the Born-Oppenheimer approximation can be omitted.[37] For each scattering angle, $\theta$, one can calculate the energy transferred, $T$, from the ion to the atom:

$$T(E_C, p) = 4 E_C \frac{M_1}{M_1 + M_2} \sin\left(\frac{\theta(p)}{2}\right)^2 \qquad (2)$$

A carbon atom is removed from the graphene lattice if the transferred energy $T(E_C, p)$ exceeds the lattice displacement energy, $E_L$. Previously reported $E_L$ values for graphene range from experimentally determined 22 eV [38] to density-functional-theory predicted 23 eV.[39, 40] We use this energy cut-off to calculate a theoretical sputter yield for ion bombardment of graphene. Defining a graphene unit cell (**Figure 1d**) with the superimposed transferred energy landscape around each carbon atom we calculated the area fraction corresponding to $T(E_C,p) \geq E_L$ (yellow area) in which incident ions transfer energy higher than $E_L$ to one carbon atom (**Figure 1e**) for creating single vacancies, $\gamma_s$. Moreover, an impact of an ion could produce a double vacancy, if the ion hits the unit cell in the area fraction, $\gamma_d$, where the transferred energy to dual carbon atoms is higher than $E_L$ (red area **Figure 1e**). The upper bound of the theoretical sputter yield, $\gamma_U$, of a defect-free, relaxed graphene lattice can be calculated by $\gamma_U = \gamma_s + 2\gamma_d$. Therefore, $\gamma_U$ corresponds to the expectation value out of the discrete probability distribution of the following three events: ion passing without sputtering; producing a single vacancy defect; and producing a double vacancy defect. The calculated upper bound corresponds therefore to the chance of C atom removal from a pristine monolayer graphene target and is necessarily higher than the experimental sputter yield which is an average removal rate in the course of graphene etching (**Figure 2a**). Continued exposure to ions removes carbon atoms from the lattice, leading to lowered probability of bombarding carbon atoms by the next ion incidence, whose effect is manifested by a decrease in the sputter yield. Therefore, it is reasonable to define a lower bound of the theoretical sputter yield, $\gamma_L$, assuming an average probability of hitting a carbon atom accounting for already created vacancies through which ions may just pass the graphene lattice with no collision. Based on this assumption we obtain $\gamma_L = \gamma_U(n_C+1)/2n_C$ (see Supporting Information), where $n_C$ denotes the total number of ions directed toward graphene. For large values of $n_C$, $\gamma_L = \gamma_U/2$.

These two bounds set for the sputter yield can bracket our measured values greatly (**Figure 2a**), suggesting that the measured sputter yield indeed corresponds to the sputtering events of clean monolayer graphene. The deviation of the measured yield from the upper bound is attributed to the gradual C atom removal from the target over the course of the ion exposure. In contrast the theoretical calculation of the lower bound does not include secondary C atom removal events such as potential removal of larger amorphous C atom agglomerates in the final phase of graphene etching. Furthermore, the excellent agreement within the employed energy range substantiates the validity of this collision theory in predicting sputter yields of graphene at various FIB conditions as well as in drawing a mechanistic explanation of the $\gamma$ values we obtained.

Since the $Ga^+$ ions carries more charges in the nucleus than $He^+$, $V(r)$ with a carbon atom can be stronger and extend wider in space. From eq. (1) and (2), this strongly repelling interatomic potential leads to large scattering angles close to the backscatter condition, $\theta = \pi$, likely transferring a substantial amount of energy to an atom in the target lattice to chop it off. Besides capturing the different aspects of $He^+$ and $Ga^+$ sputtering, our model provides an accurate description about decreasing sputter yield of monolayer graphene with increasing kinetic energy of incident energetic ions (**Figure 2a**), in good agreement with previous predictions.[17, 18] At lower kinetic energies the approaching ions get slower, interaction time prolongs, and the resultant scattering-angle distribution would become wider. Specifically, the interaction cross section extends wider in space, and since the average kinetic energy of the ions is still at least two-order-of-magnitude higher than $E_L$, the bombarded atom in the lattice could possibly be removed upon, the results indicate that the repulsive interaction of the defect-free monolayer graphene becomes very strong when the energy of the colliding particle is comparable to or lower than the lattice displacement energy, corroborating the barrier property of monolayer graphene previously reported.[6] The good model fit, indicates that indeed the interactions between the ion and the graphene can be modeled as binary collisions where single C atoms are removed from the lattice when hit by the incoming ion. Therefore, previous theoretical findings are confirmed which show that in the energy range studied the indeed create only single or double vacancy defects.[18] Only significantly lower energies <200 eV or >50 keV would lead to other effects like C atom substitution / ion implantation[28] and graphene amorphization events in the vicinity of the impact position,[18] respectively. Furthermore, the nature of the experimental design and the simplified modeling blur the effects of precise impact position on the C atom removal rate but gives a statistical average over all possible positions therefore corresponds to the realistic situation during graphene etching by energetic ions.

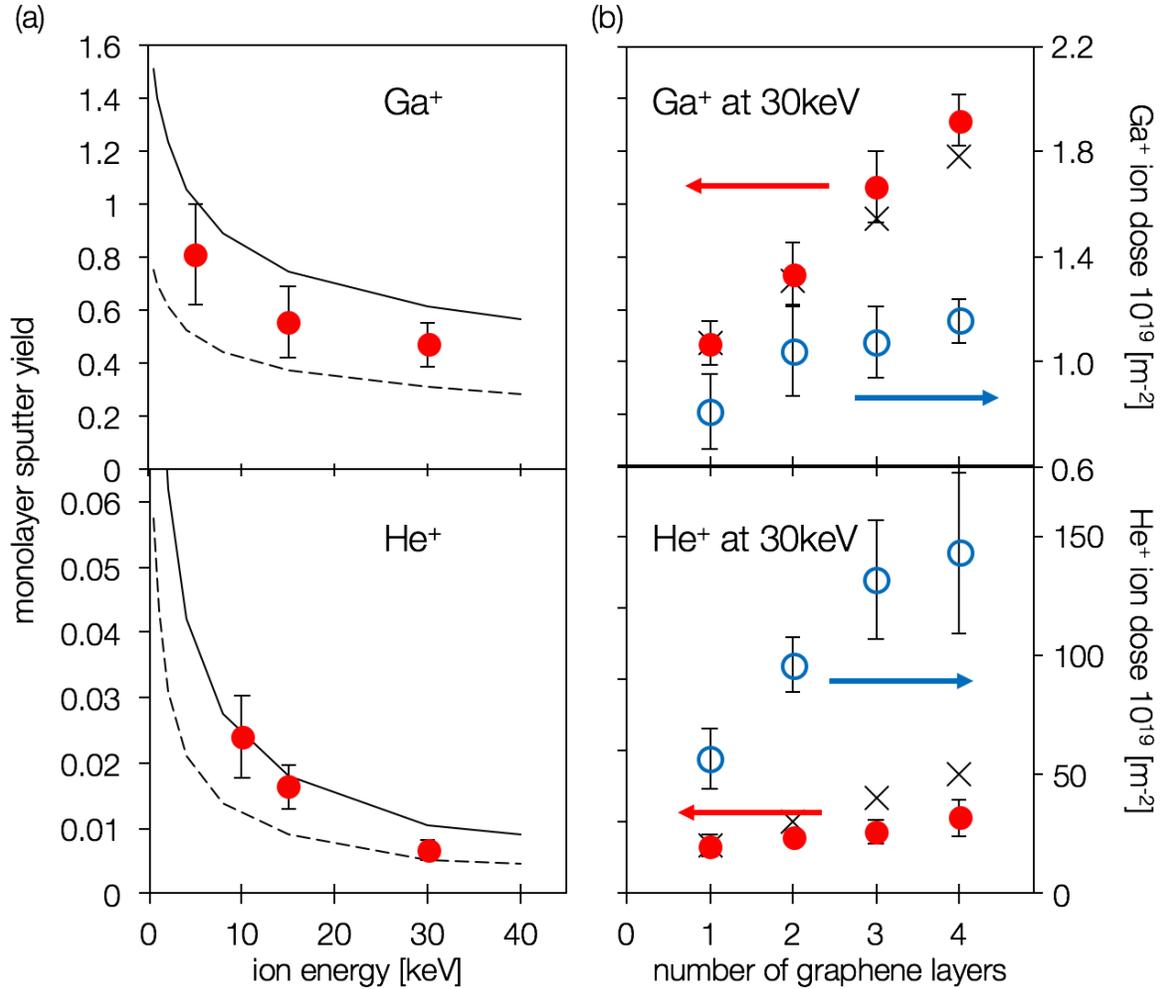

**Figure 2.** (**a**) Energy dependent, monolayer graphene sputter yields. Lower and upper bound of theoretical sputter yield for Ga$^+$ and He$^+$ using binary collision model dashed black line and solid black line, respectively. Red dots are measured sputter yield for Ga$^+$ and He$^+$ ions, with error bars indicating the standard deviation. Each data point is calculated as a mean of more than 20 independent measurements. (**b**) Layer number dependent sputter yield and ion dose necessary to pattern graphene using 30 keV Ga$^+$ and He$^+$ ions. Red dots show sputter yield for 1-4 freestanding layers of graphene, and error bars indicate the standard deviation. Black crosses indicate the expected sputter yield assuming the layer number independent probability of the carbon atom removal upon collision. Blue circle shows the experimentally measured ion dose for pattern formation.

For multilayer graphene samples we observe an increase of the sputter yield (**Figure 2b**). As a result the ion dose necessary to etch a pattern into freestanding multilayer graphene does not linearly increase with the number of layers (**Figure 2b**) eg. a 4 layer graphene sample requires only 1.2 × $10^{19}$ $m^{-2}$ $Ga^+$ ions at 30 keV which roughly twice the dose required for a monolayer graphene. For $He^+$ ions the effect is less pronounced where 3 times the dose of monolayer patterning is required to create a pattern in a 4 layer graphene sample (**Figure 2b**). In general results for freestanding multilayer graphene etching are approaching the sputter yield reported for bulk carbon allotropes.[32, 33] In sputtering bulk materials the so-called primary event – the collision of the incident ion with the target atom – does not play a crucial role in target atom removal. Though, the secondary sputtering events – the collisional cascade inside the target material leading to a momentum inversion – contribute mainly to the bulk sputter yield. The events responsible for the target atom removal in the 3D materials are the following: first, cascade of vibration energy into the target material could excite neighboring atoms in the bulk lattice; and second, these neighboring atoms recoil causing a local inversion of the momentum and could escape the bulk if they were close to the surface and ended up gaining sufficient energy. These events could often take place when the bulk material thicknesses are ~20 nm and ~5 nm for $Ga^+$ and $He^+$, respectively. However, in the present case even the multilayer graphene samples are comparably thin having maximum quadruple layers. Therefore, we intuitively surmise that only the probability of ions hitting a carbon atom while passing through the multiple graphene layers should increase with number of layers. This case can be approximated by modeling the collision with each graphene layer as independent stochastic event with constant probability of collision in each layer (assuming that the kinetic energy of a scattered ion remains intact upon collision in a layer). On the other hand, the removal of carbon atoms in one layer would lower the probability that newly incident ions interact with atoms of the same layer since the number of layers at this particular spot is reduced by one. Therefore, the average expected sputter yield $\gamma_N$ for an $N$-layer-thick sample equals to $\gamma_N = \gamma_{ion} (N + 1)/2$. Using this assumption, we calculated an increase in sputter yields with layer number for both He$^+$ and Ga$^+$ FIB processes (**Figure 2b**). For freestanding multilayer graphene, the experimentally observed sputter yield for Ga$^+$ ions increases from ~0.5 for monolayer graphene to ~1.3 for quadruple layer samples matching nicely the theoretically prediction (**Figure 2b**). The consistency between experiment and theoretical prediction shows that interestingly the overall escape of the C atom from the multilayer graphene samples is not inhibited. C atom removal in multilayer graphene could follow the route of a cascading collision of equal collision partners. Once a C atom in the first layers is hit it recoils and collides with a C atom in the lower layers from where the cascade continues until a C atom in the last graphene layer is removed by forward sputtering or even more complicated effects like catalytic etching in the presence of an underlying graphene layer [38] may cause this counterintuitive finding. For He$^+$ ion the observed increase in sputter yield is less than the model prediction (**Figure 2b**), possibly attributed to a less efficient vibrational cascade since the average transferred energy to the target atom is significantly lower than in the case of Ga$^+$ ion sputtering (see Supporting Information). Both results show that in contrast to monolayer graphene ion interaction the presence of additional graphene layers require more extensive modeling attempts.

Vacancy generation mechanism in graphene under energetic ion bombardment can be further elucidated by Raman spectroscopic monitoring of the evolution of pristine freestanding graphene subject to various ion irradiation doses. 2D Raman maps are used to acquire a representative Raman signal from the pristine and irradiated graphene (**Figure 3a**). The Raman spectrum of pristine graphene verifies high quality monolayer with a sharp G peak of FWHM 23.5 cm$^{-1}$ at ~1587 cm$^{-1}$ (**Figure 3b**), indicative of abundance of sp$^2$-hybridized carbon atoms rather than sp$^3$.[41] A sharp second harmonic benzene breathing mode commonly named a G' peak around ~2679 cm$^{-1}$ shows a single peak of twice the intensity of the G peak (**Figure 3b**), indicative of defect-free monolayer graphene.[42] Defects in the graphene lattice can affect the intensity of the D band (at ~1340 cm$^{-1}$) in the Raman spectrum. This first harmonic or the radial breathing peak of the benzene ring unit activated in the case of asymmetry in a sp$^2$-hybridized lattice[41] can originate largely from the vicinity of grain boundaries or vacancy sites and serves as a convenient measure of defect densities of any kind. In particular, the intensity ratio of the D and G peaks, I(D)/I(G), has been shown to follow a characteristic dependency on the defect density.[43] Using the experimentally determined $\gamma$ and the areal dose, $\sigma$, of applied ions we could calculate the average defect distance, $L_D = (\gamma_{ion}\sigma_{ion})^{-1/2}$. The obtained result for the I(D)/I(G) with respect to $L_D$ for the He$^+$ FIB nicely matches an empirical equation previously reported[43] about low energy Ar$^+$ bombardment (**Figure 3c**). Indeed, this agreement corroborates our finding that a very high portion (>99% at 30 keV) of He$^+$ is passing through graphene without generating any lattice vacancy. It is the rare collision events that produce a single vacancy defect on graphene.

The non-monotonic relation of I(D)/I(G) over the defect distance for freestanding graphene stands in contrast to the reported Tuinstra-Koenig relation reported for graphite.[44] Unlike bulk graphite where ion irradiation renders the surface to sp$^3$-bonded amorphous carbon yielding a monatomic increase of I(D)/I(G), ion irradiation of graphene exhibits three distinct regimes of etching.[43] In the first regime (large $L_D$), single vacancies are produced which lead to increased lattice disorder and a strong D peak. The total number of sp$^2$ bonds remains nearly constant (G peak), yielding an increase of the I(D)/I(G) ratio (**Figure 3c, I**). In the second regime ($L_D <$ ~4 nm), the continuous removal of carbon atoms from the graphene layer destroys the hexagonal lattice, and the radial breathing (D peak) of benzene rings decreases in the abundance of intact benzene rings. Still, the layer comprises sp$^2$-hybridized amorphous carbon chains accounting for the G peak. In this regime I(D)/I(G) is getting smaller again (**Figure 3c, II**). In the third regime ($L_D < \sim 1$ nm), the lose network of sp$^2$-bonded carbon atoms is removed without significant changes in the bonding structure therefore the I(D)/I(G) remains constant around the unity with total peak intensities vanishing slowly until nearly all the atoms are removed (**Figure 3c, III**).

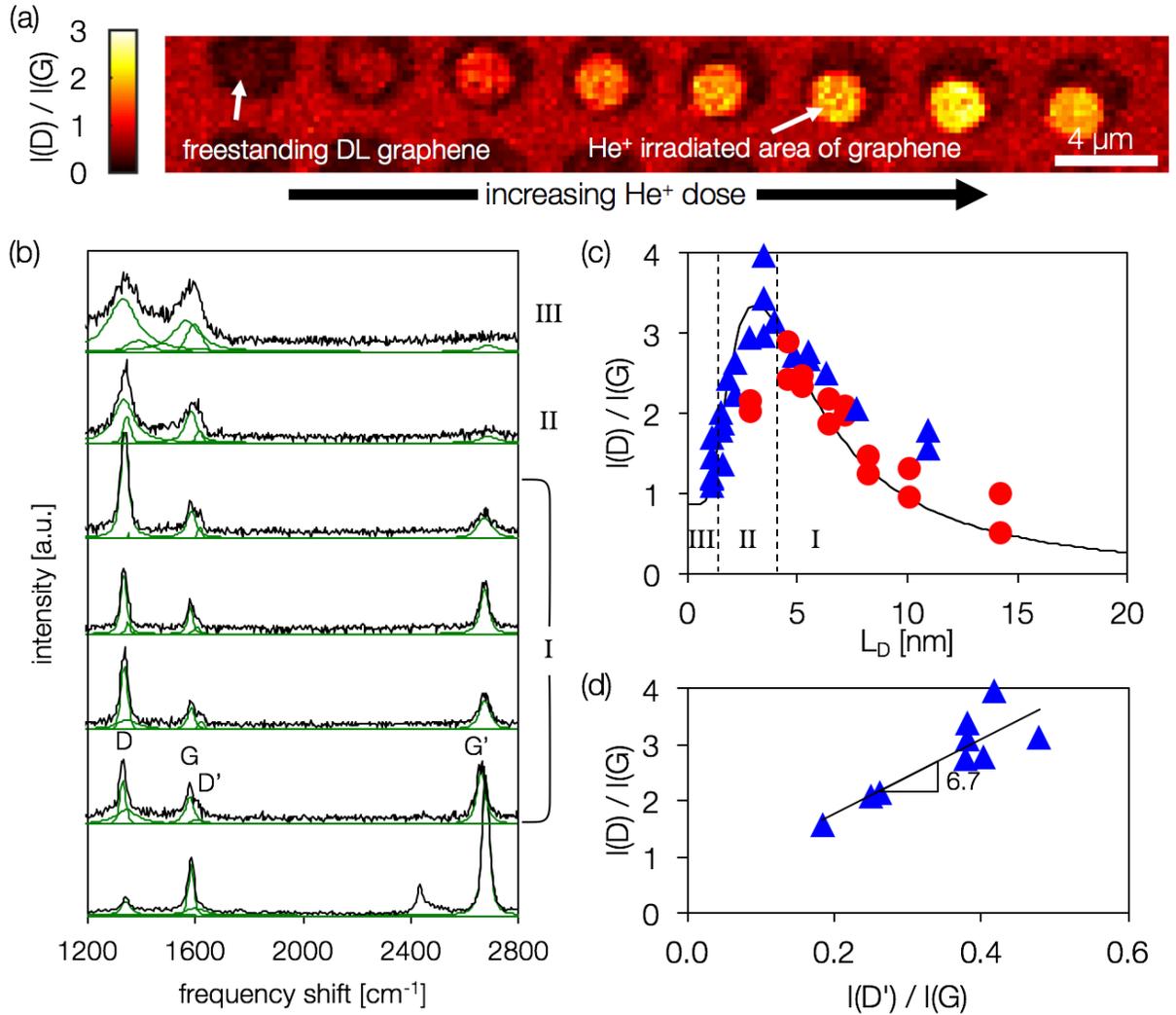

**Figure 3.** Raman analysis of He$^+$ ion irradiated graphene. (**a**) 2D Raman map of I(D)/I(G) on freestanding double layer graphene with different exposure dose of He$^+$ ion dose ranging from 0, 2.5, 3.8, 5.0, 6.3, 9.4, 12.5, 31.3 × 10$^{18}$ m$^{-2}$ (**b**) Raman spectra of irradiated monolayer graphene with ion doses of 0, 1.25, 2.5, 5, 12.5, 56, 125 ×10$^{18}$ He$^+$/m$^2$ (from bottom to top). (**c**) Measured I(D)/I(G) ratio versus average defect distance $L_D$ for monolayer graphene (blue triangle) and double layer graphene (red circle) as compared with calculated empirical function (solid black line) from reference [43]. (**d**) I(D)/I(G) vs I(D')/I(G) for monolayer graphene exposed to low He$^+$ ion doses (1.25-62.5 × 10$^{18}$ m$^{-2}$), showing linear proportionality (solid black line).

The creation of individual vacancy defects in the regime I during the He$^+$ FIB of graphene can be supported by spectral decomposition of Raman spectra. The evolution of D (~1340 cm$^{-1}$) versus D' peaks (~1620 cm$^{-1}$) for low $\sigma_{ion}$ follows a linear proportionality of I(D)/I(D') ≈ 7 (**Figure 3d**) in great agreement with a previous report.[45] This spectral behavior clearly differs from the other defect creation mechanisms such as sp$^3$ bond creation or grain boundaries, which lead to I(D)/I(D') ≈ 13 and I(D)/I(D') ≈ 3.5, respectively.[45, 46] Therefore, albeit the low collision probability, irradiation of a low dose of energetic He$^+$ ions onto freestanding graphene could instigate a single ion graphene interaction facilitating single-vacancy-type defect generation in the lattice, preferentially. Furthermore, the resultant I(D)/I(D') evolution shows clearly that the Raman spectrum does not arise from sp$^3$ amorphous carbon deposit or contamination. Therefore, the precise control of $\sigma_{He^+}$ at a given He$^+$ energy promises to create array patterns of vacancy defects.

The evolution of the I(D)/I(G) ratio of double layer graphene follows the same trend as the monolayer graphene (solid red circles, **Figure 3c**). When calculating the average distance ($L_D$) in which defects are created by considering the area of both graphene layers $L_D = N^{1/2}(\gamma_{ion}\sigma_{ion})^{-1/2}$ (with $N=2$), the results for monolayer and double layer graphene coincide each other, which corresponds to the case where the same amount of ions directed toward the graphene is distributed to twice the amount of lattice atoms. Since the evolution of the I(D)/I(G) ratio is the same as on monolayer, it indicates that the vacancy formation by energetic He$^+$ ions on double layer graphene is independent of the presence of a second layer. Therefore, we can confirm that each He$^+$ ion hitting the graphene either sputters one of the graphene layers or penetrates without creating a vacancy, and sputtering caused by the vibration energy cascade is nearly negligible.

Analysis of the Raman spectrum of ion-irradiated freestanding graphene elucidates the pattern formation on freestanding graphene via FIB (**Figure 4**). It follows the route of gradual vacancy formation at the initial step ensued by defect agglomeration that ends up amorphizing on the ion-beam irradiated area of the graphene lattice as recently proposed by TEM study of irradiated graphene.[29] At the last step of patterning the amorphous yet $sp^2$-hybridized carbon layer is etched away.

Using these insights we could, for the first time, achieve the smallest pore-array patterns on graphene perforated by FIB. The 2D nature of graphene prefers forward sputtering such that each particle removal from the graphene lattice is caused by the particle collision with an incident energetic ion. On freestanding double layer graphene we cut holes with ultimate precision and repeatability at relatively high rates allowing an efficient large scale pattern formation. With ~$10^4$ $Ga^+$ ions per pore at 30 keV in a single-pixel exposure experiment, we could drill into double layer graphene an array of holes smaller than 9 nm (with a mean diameter of 5.8 nm) at the average spacing of 50 nm (**Figure 4a**). Reducing the $Ga^+$ ion dose on monolayer graphene to 2500 $Ga^+$ ions per pore we obtained pore sizes of 3.5 nm (**Figure 4c**), which is significantly smaller than previously reported results of sub-10-nm pores on graphene.[25] The tight control of the exposed ions does not only allow to pattern at the resolution limit of the Ga FIB system, which is defined by the beam size (~4 nm). The low dose guaranties an exceptional short process time of few us per pattern, enabling even large scale patterning of graphene for device integration.

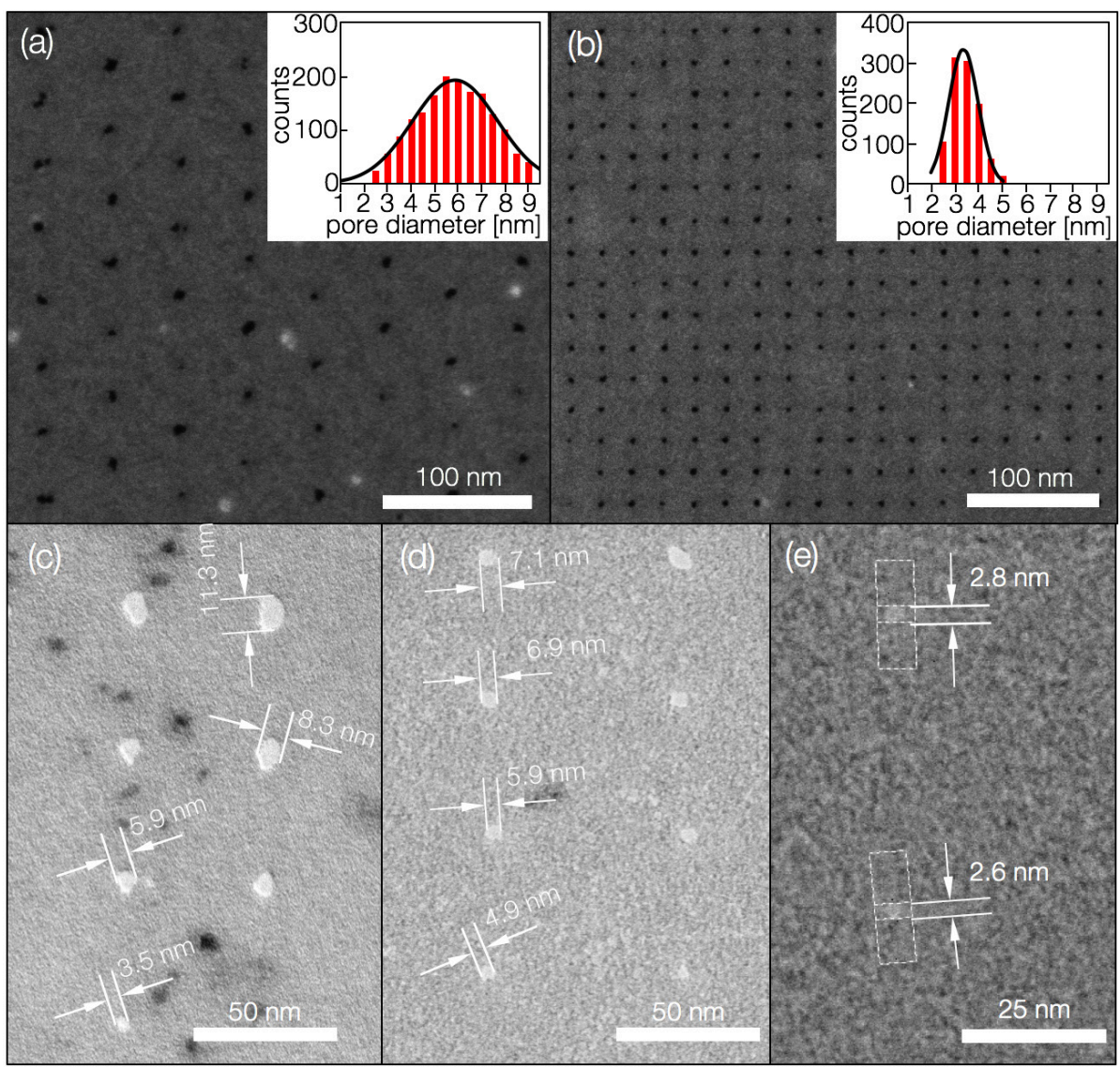

**Figure 4.** Pore arrays on graphene perforated via FIB milling. Secondary-electron-detection-mode SEM images of pore arrays having 50-nm-wide (**a**) and 25-nm-wide (**b**) pitches on freestanding double layer graphene perforated by 30-keV Ga$^+$ at ~10$^4$ per pore (**a**) and by 30-keV He$^+$ at ~4.4×10$^5$ per pore (**b**), respectively. Dark contrast indicates pores. The resultant pore-size distributions are 5.9±3.6 nm (**a**) and very narrow 3.3±1.2 nm (**b**). (**c**) Transmission-electron-mode (bright field) SEM image of pores on freestanding monolayer graphene perforated by 30-keV Ga$^+$ at 2500-10$^4$ per pore (bottom to top), exhibiting pore sizes from 3.5 to 11.3 nm. (**d, e**) Transmission-electron-mode (bright field) SEM images of pores on the monolayer graphene perforated by 30-keV He$^+$ at 3.9-6.2×10$^5$ per pore (**d**, bottom to top) and 2.7×10$^5$ He$^+$ per pore (**e**), showing pore sizes of 4.9-7.1 nm (**d**) and 2.6 and 2.8 nm (**e**), respectively.

Note that these top-down-drilled pore sizes are significantly smaller than sputtering of bulk material could produce. In 3D sputtering process, an incident ion collides with multiple target atoms to initiate a collision cascade within the target material. Recoiling target atoms can induce a momentum inversion of certain atoms close to the surface, resulting in an escape from the bulk phase.[19] These secondary events occur in the vicinity of the incident ion spot called an interaction diameter of which reported values are around 20-30 nm.[47, 48]

For ion exposure of graphene in HIM we achieved even smaller feature sizes. Here again we could overcome the previously reported interaction diameter limit of 5 nm[47] and repeatedly patterned pore arrays into freestanding double layer graphene with mean diameter of 3.4 nm at 25-nm-wide spacing using 30-keV $He^+$ ions at $4.4 \times 10^5$ per pore (**Figure 4b**). The effect of total ion dose on the pore size can be observed by exposing a freestanding monolayer graphene with $3.9 \times 10^5$ to $6.2 \times 10^5$ $He^+$ ions (**Figure 4d**). Despite the single-pixel exposure we see a pore size increase from 4.9 to 7.1 nm caused by the imperfect spot shape of the irradiating beam. By decreasing the number of $He^+$ hitting the monolayer graphene to $2.7 \times 10^5$ we were able to produce holes with 2.6-nm-wide diameter (**Figure 4e**), comparable to pore sizes drilled in graphene using TEM systems.[21, 49] Our results show a significant advancement in graphene patterning via FIB milling in terms of feature size and array dimension, enabled by detailed knowledge of the interaction mechanisms involved and the 2D nature of our target material. Use of freestanding graphene allowed us to create patterns while avoiding undesirable secondary effects during FIB milling (e.g., ion implanting and substrate swelling), unlike frequently reported for the supported graphene samples.[13, 15, 32, 50]

CONCLUSION:

Experimental and theoretical investigations of the interaction between freestanding graphene layers and energetic ion irradiation confirm that pristine graphene could be transparent to material at elevated kinetic energy suggested by previous theoretical investigations.[17] For instance, graphene is highly transparent to 30-keV-accelerated $He^+$ particles, only ~1% of which collide with the graphene lattice and sputter carbon atoms as compared with 47% for $Ga^+$ (30 keV). Both binary collision theory and experimental characterizations point out the uniqueness of the 2D material sputtering in that the major sputtering mechanism would be a simple binary collision between incident ion particle and carbon atom in the lattice, in clear contrast to vibration energy cascade and recoiling for sputtering 3D material. The sputtering probability (i.e., sputter yield) of the freestanding graphene layers depends strongly on the species and kinetic energies of the bombarding ion particles, as well as on the number of the layers. These findings lead to a rationale for patterning the 2D material such that precise control of the local ion exposure dose on freestanding 2D materials could bring the size limit of this technology to a new level. This pattern generation strategy was firstly attested by creation of hole-array patterns with feature sizes down to 2.6 nm and 3.5 nm using $He^+$ FIB and Ga+ FIB, respectively. These results highlight the great potential of efficient, sub-5-nm-scale feature generation on 2D materials, opening up great possibilities in the nanomanufacturing of devices that employ flexible 2D materials patterned at nanometer scales.

MATERIALS/METHODS:

Freestanding graphene sample preparation:

Polycrystalline graphene was synthesized via CVD using conditions optimized for continued monolayer coverage reported elsewhere.[51] A Cu foil (Alfa Aesar 46986) is cleaned by Ar ion beam milling (10 min at 250 mA, 600 V acceleration) prior to a reduction annealing in a $H_2$-rich ambient (100 sccm $H_2$ in 1500 sccm Ar) at 950 °C for 30 min. Growth is initiated by addition of 25 sccm $C_2H_4$ to the chamber for 2 min and subsequent 50 sccm for 1 min. As-grown monolayer graphene was transferred onto a punctured $SiN_x$ membrane supported by a Si chip (destination substrate) using a wet transfer method.[7] PMMA is spun onto of the graphene-coated Cu foil. Cu is etched away in $(NH_4)_2S_2O_8$ (0.5 M) leaving the graphene/PMMA film afloat the etchant. For multilayer graphene samples, this graphene/PMMA film is fished after rinsing in DI water with another graphene-grown Cu foil, and the additional Cu substrate is etched again. Repeating this step until the desired number of layers is reached a clean multilayer graphene sample without interlayer contamination is obtained. After final rinsing in DI water the floating graphene/PMMA film is fished with the porous $SiN_x$ destination substrate. Upon drying the graphene/PMMA film, the graphene is cleaned using well-known thermal decomposition methods[52, 53] where PMMA is pyrolized away at 400°C in 900 sccm $H_2$ and 100 sccm Ar, yielding clean freestanding graphene.

Graphene characterization:

The quality and cleanliness of the graphene samples was confirmed using scanning electron microscopy (Helios 450, FEI). Furthermore, the transferred graphene was analyzed before and after the patterning by 2D Raman mapping (micro Raman CRM200, WiTec) using a 532-nm incident laser beam at 0.4 mW with a pixel spacing of 100-250 nm.

Ion irradiation:

The freestanding graphene layers were irradiated with ions at different acceleration voltages via FIB (Helios 450, FEI). Toward the graphene sample $Ga^+$ ions are accelerated at 5-30 keV with probe currents ranging from 1.1 pA to 40 pA and a chamber pressure of $\sim 7\times 10^{-5}$ Pa. For the $He^+$ ion irradiation we used a He ion microscope (Zeiss Orion) equipped with a pattern generator (Raith Elphy MultiBeam) operated at 10-30 keV using a probe current of 5-17.5 pA at a chamber pressure of $\sim 7\times 10^{-5}$ Pa. Note that relatively high probe currents are desirable to reduce ion-beam-induced deposition in the exposed areas (Supporting Information). On both tools the ion dose was controlled by the exposure dwell time of each pixel ranging from 100 ns to 8 ms.

Experimental sputter yield determination:

The sputter yield ($\gamma$) defined as an average removal rate of carbon atoms from the graphene lattice was determined by the following experiment. Circular pattern arrays of ions are accelerated toward freestanding graphene with increasing areal doses ($\sigma$). On each circular pattern the energetic ions are evenly distributed with 200 nm in diameter (nominal area $A_n = 0.0314$ µm²). After the patterning we acquire SEM micrographs and measure the resulting pattern area $A_p$ using an image analysis program (ImageJ). The lower bound number of removed carbon atoms $N_C = \sigma_C A_p N$ can be easily calculated by using the areal density of carbon in graphene $\sigma_C$ (m⁻²) and the number of graphene layers, $N$. The sputter yield is defined as $\gamma_{ion} = N_C/N_{ion}$, where $N_{ion} = \sigma_{ion} A_n$ is the total number of ions irradiating the graphene layer.

Graphene pore characterization:

Sizes of pore arrays in double layer graphene were determined using high-resolution SEM images obtained on a FEI Helios 450 at 5 keV, 13pA probe current collecting secondary electrons. The micrograph was analyzed (ImageJ) with identifying the circular pore area by the dark regions (no secondary electrons generated), from which the diameter or the pore dimension was calculated. Electron micrographs of the smallest pores created in monolayer graphene were obtained using Hitachi SU8230 SEM at 30 keV and 55 pA probe current and equipped with an SEM detector for a bright-field transmission electron mode (aperture size: 1 mm). Pore area appears as bright area where electrons pass the sample without being scattered. The micrographs were analyzed using the Gatan Digital Micrograph image analysis software.

AUTHOR INFORMATION

**Corresponding Author**

\* To whom correspondence should be addressed. E-mail: E-mail: ivan.shorubalko@empa.ch (IS), parkh@ethz.ch (HGP).

**Author Contributions**

JB, IS and HGP conceived the study and designed the experiment. RW provided the freestanding graphene samples. JB and IS performed the experiments. JB analyzed the data. The manuscript was written through contributions of all authors. All authors have given approval to the final version of the manuscript.

ACKNOWLEDGMENT

We appreciate the support from Binnig and Rohrer Nanotechnology Center of ETH Zurich and IBM Zurich. J.B. thanks Swiss National Science Foundation for financial support (200021-137964). I.S. is grateful to Swiss National Science Foundation for support in equipment procurement (REquip 206021-133823). This work was partially supported by LG Electronics Advanced Research Institute, for which H.G.P. and R.M.W. are thankful. I.S. appreciates Y. Zheng at EMPA (Swiss Federal Laboratories for Materials Science and Technology), Switzerland, for his initial contribution on the perforated graphene imaging.

SUPPORTING INFORMATION

**Transferred Energy**

The transferred energy to the target atom in the graphene strongly depends on the collision parameter p. Therefore, one can calculate the amount of energy transferred to the C atom depending on the area fraction $\gamma_A = p^2\pi/A_{graphene}$ of the graphene unit cell covered by p (**Figure S1**). Since energetic ions hit the graphene unit cell at a random location, the area fraction can be identified with the fraction of C atoms hit the graphene target. As a result, one can estimate the percentage of C atoms gaining a certain transferred energy $T$. As an example ~50% of Ga$^+$ ions gaining more than the lattice displacement energy have energies higher than ~200eV compared to only ~15% in the case of He$^+$ (**Figure S1**).

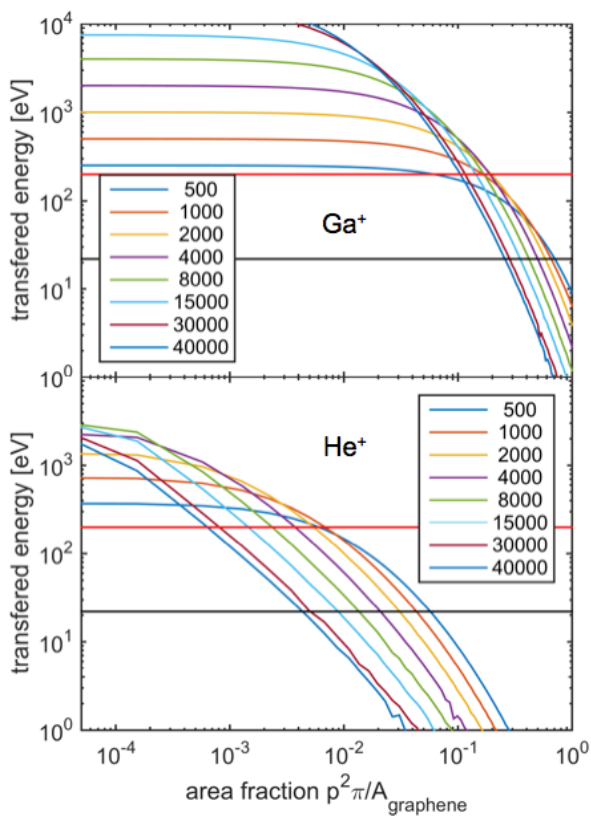

**Figure S1.** Transferred energy for Ga$^+$ and He$^+$ ion irradiation versus area fraction of the graphene unit covered by collision parameter p. Energy of incident ions given in eV. Two cutoff energies are displaced: black line corresponds to lattice displacement threshold of 23eV necessary to remove a C atom form the graphene lattice whereas the red solid line depicts the threshold line of 200 eV indicating the fraction of hit C atoms having significant recoiling energy.

## Lower Bound of Theoretical Sputter Yield

The lower bound of the sputter yield can be derived by the following consideration. Initially the graphene sheet contains total number of carbon atoms, $n_C$. Each carbon atom has a scattering cross section area of $A_C$ defined by the area in which the transferred energy of the ion hitting carbon atoms exceeds the lattice displacement energy $E_L$ (see main text **Figure 1e** yellow area). The total scattering-cross-section area occupied by the carbon atoms meeting the sputtering condition equals to $n_C A_C$. Initially the probability of hitting a carbon atom, $p_1$, equals to the ratio of the total scattering-cross-section area to the total defined pattern area, $A_t$, yielding $p_1 = n_C A_C / A_t$. Once sputtering occurs in this event the number of carbon atoms is reduced by one, giving a new probability, $p_2 = (n_C - 1) A_C / A_t$, to the next carbon atom to be removed from the 2D lattice. By continuing this argument all the way to the rest of carbon atoms in the lattice, an average sputtering probability is obtained:

$$\bar{p} = \frac{n_C A_C / A_t + (n_C - 1) A_C / A_t + (n_C - 2) A_C / A_t + \cdots + (n_C - n_C + 1) A_C / A_t +}{n_C} = \frac{\sum_{i=1}^{n_C} i A_C / A_t}{n_C}.$$

Now noticing that $A_C / A_t$ is constant and that $A_t$ can be expressed in number of initial carbon atoms $n_C$ and unit cell area $A_u$ ($A_t = 1/2 \, n_C A_u$), one sees that $A_C / A_t = 1/n_C \, \gamma_U$ contains the upper bound of sputter yield, $\gamma_U$. Therefore the mean probability can be rewritten as $\bar{p} = \frac{\sum_{i=1}^{n_C} i}{n_C^2} \gamma_U$. Noticing the summation over $i$, one immediately arrives at $\gamma_L = \frac{n_C + 1}{2 n_C} \gamma_U$.

## Limiting Ion Beam induced deposition

On the He⁺ FIB we used relatively high probe currents of 5-17 pA. These values are significantly higher than the standard imaging conditions and previously reported patterning currents of 0.5-1 pA.[15] However, we found the high probe currents to be necessary for large-scale graphene patterning because they enable to pattern $10^6$ pores of sub-5 nm in size within 2 hours. For low currents we found out that dose required for the patterning increases substantially (by factor of 10-50). We attribute this increase to the deposition of amorphous carbon material around the patterned area, which prevents readily removing of carbon atoms from the graphene lattice (**Figure S2a**).

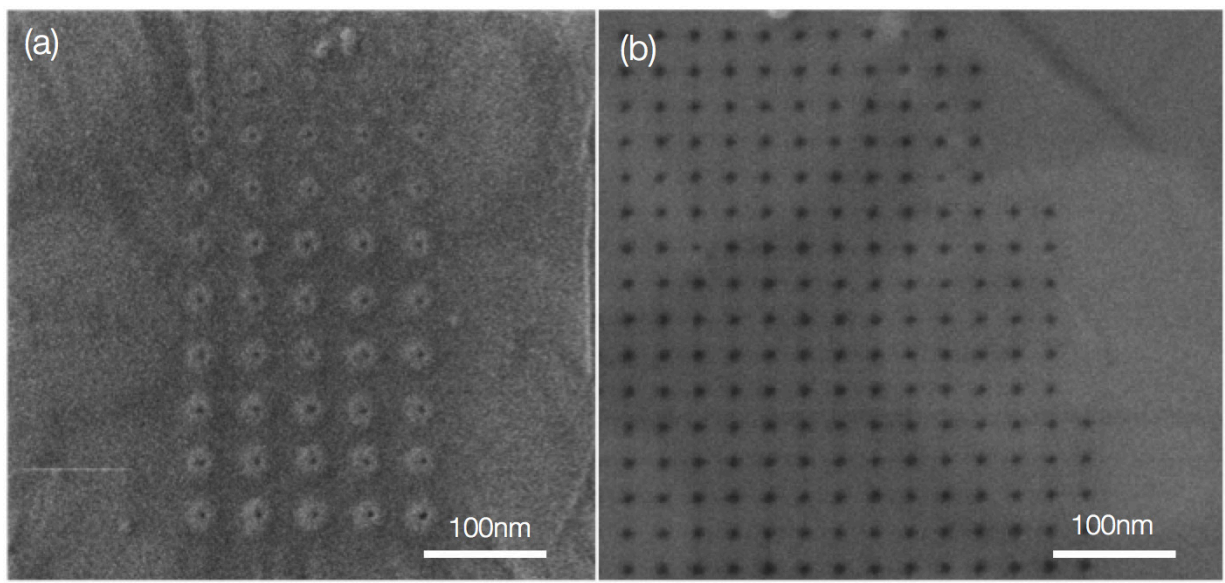

**Figure S2.** He⁺ ion micrograph of etched pores in HIM. (**a**) Probe current of 1.4 pA and dwell times from 100 ms to 4600 ms (top to bottom row in steps of 500 ms). Strong deposition of material is visible around each patterned feature. Only beam-irradiation dwell times higher than 1600 ms show pores clearly visible in the micrograph. (**b**) Probe current of 10.5 pA with a dwell time of 15 ms leads to pore formation with no deposition in the vicinity of it.

Ion beam induced deposition is a widely reported phenomenon.[54] Gas molecules (e.g., volatile carbon species in the containing chamber) can adsorb on the target substrate, or the target substrate could already been contaminated by various adsorbates. When incident ions are inelastically scattered and create secondary electrons, these electrons can collide with and dissociate the adsorbates to leave nonvolatile compounds on the surface. This chemical reaction is limited by both the supply of adsorbing molecules as well as the reaction energy provided by the incoming ion flux. At low ion fluxes the contamination molecule mobility is high enough to move to the reaction site and build up material. In this case the reaction is limited by the energy input. This mechanism can be responsible for the deposition around the defined pattern (**Figure S2a**) and the very large dose values necessary to etch a hole onto graphene. At high ion fluxes the surface diffusion of the adsorbates to the patterning site is relatively slow, shifting the reaction toward a diffusion-limited regime. Using these patterning parameters